\documentclass[final,1p,times]{elsarticle} % do not change
\usepackage{graphicx} % do not change
\usepackage{amssymb} % do not change
\usepackage{amsthm} % do not change
\usepackage{lineno} % do not change

\journal{Nuclear Physics A} % do not change
\begin{document} % do not change

\begin{frontmatter} % do not change

%% QM09Author: please enter your  
%% Title, author and address info here; please do not use footnotes

% Your Title - please insert
\title{Investigating Sources of Angular Correlations at High $p_T$
in~Nucleon--Nucleon and Nucleus--Nucleus Collisions at~the~CERN SPS}

% Principle author, and co-authors - please insert
\author{Marek Szuba$^{a}$ for the NA49 Collaboration}

% Address - please insert
\address[a]{Faculty of Physics, Warsaw University of Technology,
Koszykowa 75, 00-662 Warszawa, Poland}

\begin{abstract} % do not change

Angular correlations of high-$p_T$ hadrons can serve as a probe of
interactions of partons with the dense medium produced in high-energy
heavy-ion collisions but other effects may also be important at SPS
energies. To study the various contributions, NA49 has performed
an energy and system-size scan of two-particle azimuthal correlations
in \textit{Pb+Pb}, \textit{Si+Si} and \textit{p+p} collisions at $\sqrt{s_{NN}}$ = 17.3~GeV,
as well as central \textit{Pb+Pb} collisions at 6.3, 7.6, 8.8 and 12.3~GeV.
These results were compared to UrQMD simulations.

NA49 results show a flattened away side of $C_{2}(\Delta\phi)$ in central \textit{Pb+Pb}
(\textit{Au+Au}) collisions which depends weakly on collision energy even for low
SPS energies. This is at odds with the standard scenario of parton
energy loss. On the other hand, the near-side peak amplitude drops visibly
with decreasing collision energy, turning into a depletion below
$\sqrt{s_{NN}}$ = 8.8~GeV. UrQMD describes the correlation functions
on the away side but disagrees on the near side.

\end{abstract} % do not change

\end{frontmatter} % do not change

%% QM09: we keep linenumbers at least for initial version
%\linenumbers

\section{Introduction}
\label{sec:introduction}

In early 2008 the NA49 Collaboration presented its first results on two-particle
azimuthal correlations of non-identified charged hadrons, concluding that the flattening
of the away side of the correlation function in most central \textit{Pb+Pb} collisions
at $\sqrt{s_{NN}}$ = 17.3~GeV, as well as the observed dependence of the near-side amplitude
on the charge of trigger and associate particle, was consistent with qualitative expectations
of QGP presence in central high-energy heavy-ion interactions~\cite{SzubaQM2008}. 

Subsequently, we studied the behaviour of the two-particle azimuthal correlation
function in central collisions on other observables: system size and collision energy.
With deconfinement in heavy-ion collisions believed to set in at low SPS
energies~\cite{NA49onset}, observation of evolution of the correlation function may
shed light on the mechanism responsible for the away-side flattening. For central collisions
subtraction of flow is not necessary, an advantage in view of the criticism of the commonly
employed subtraction techniques~\cite{gyulassyZYAM, trainorZYAM}.

The present analysis is based on the following data sets of NA49: central \textit{Pb+Pb} collisions,
$\sigma/\sigma_{geom}$ = 0--5~\%, at $\sqrt{s_{NN}}$ = 17.3, 12.3, 8.8, 7.6 and 6.3~GeV; central
\textit{Si+Si} collisions, $\sigma/\sigma_{geom}$ = 0--5~\%, at $\sqrt{s_{NN}}$ = 17.3~GeV;
\textit{p+p} collisions ($\approx$90~\% inelastic) at $\sqrt{s_{NN}}$ = 17.3~GeV.

We have compared the experimental correlation functions to results from simulated events using
the string-hadronic model UrQMD~2.3~\cite{urqmd1, urqmd2}, which allows optional incorporation
of jet production from PYTHIA~\cite{pythia}.

\subsection*{The Method}

The method of calculating two-particle azimuthal correlation functions was described
in detail in our previous report~\cite{SzubaQM2008}. Acceptance-corrected correlation functions $C_{2}(\Delta\phi)$
were obtained in the $\Delta\phi$ range of $\left[ 0,~\pi \right]$. The transverse momentum selections
remain unchanged: $2.5~GeV/c \le p_T^{trg} \le 4.0~GeV/c$ for trigger particles
and $1.0~GeV/c \le p_T^{asc} \le 2.5~GeV/c$ for associates.

Additionally, two new techniques have been introduced. The ``central \textit{Pb+Pb} at $\sqrt{s_{NN}}$
= 17.3~GeV'' correlation function used as reference for other functions in the scan was
parametrised with a two-part polynomial fit (third-order on the near side, linear on the away side). This
has greatly improved the ease of comparisons. Secondly, each correlation function
from the energy scan has been fitted with two linear functions (one for the near side, one for the away side),
substituting the comparison of peak values by comparing the slopes of lines fitted to $C_{2}(\Delta\phi)$.

For all correlation functions statistical errors are plotted as bars, with systematic uncertainties illustrated
using gray boxes. In case of values extracted from fits, their error bars combine statistical and systematic uncertainties.

\section{Results}
\label{sec:results}

\subsection{System-size Scan}
\label{sec:systemSizeScan}

\begin{figure}[htb]
  \begin{center}
    \includegraphics[width=0.49\textwidth]{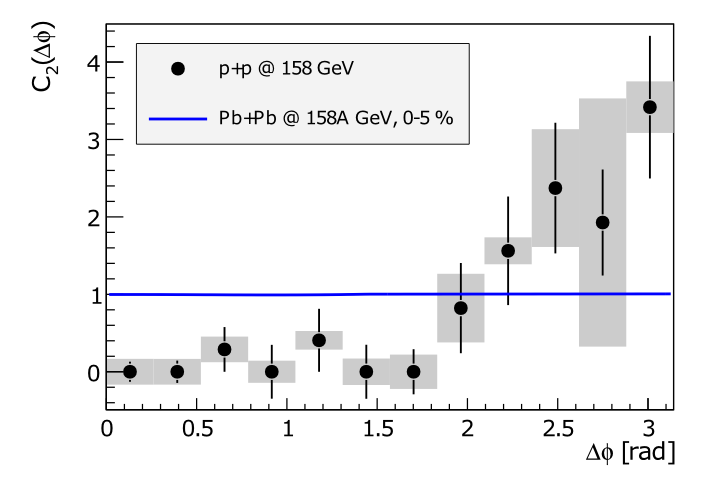}
    \includegraphics[width=0.49\textwidth]{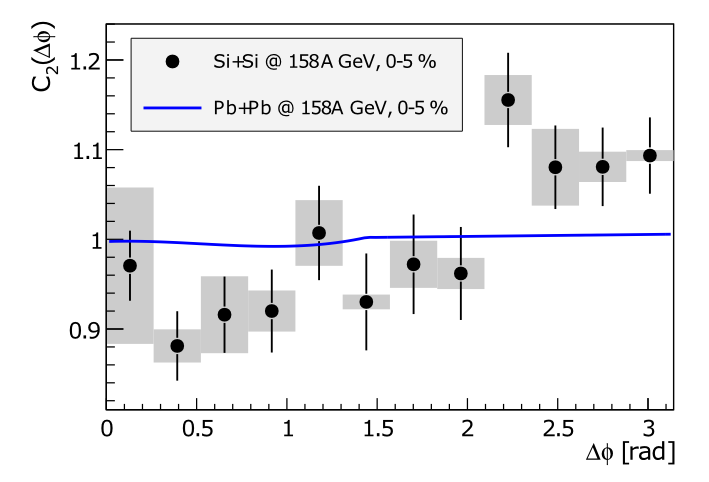}
  \end{center}
  \caption{Two-particle correlation functions from \textit{p+p} (left) and central \textit{Si+Si} (right) events
    at $\sqrt{s_{NN}}$ = 17.3~GeV, compared to a parametrisation of central-\textit{Pb+Pb} results at the same energy (curves).}
  \label{fig:systemSizeScan}
\end{figure}

Figure~\ref{fig:systemSizeScan} shows two-particle azimuthal correlation functions from \textit{p+p}
and central \textit{Si+Si} collisions at $\sqrt{s_{NN}}$ = 17.3~GeV, compared to a parametrisation of
\textit{Pb+Pb} results~\cite{SzubaQM2008}. Overall strength of the correlation becomes significantly larger
as the system size decreases. Moreover, no flattening of the away-side peak, as present in central heavy-ion events,
is visible in \textit{Si+Si} and \textit{p+p} collisions --- indeed, the peak becomes narrower
with decreasing system size.

\subsection{Energy Scan}
\label{sec:energyScan}

\begin{figure}[htb]
  \begin{center}
    \includegraphics[width=0.9\textwidth]{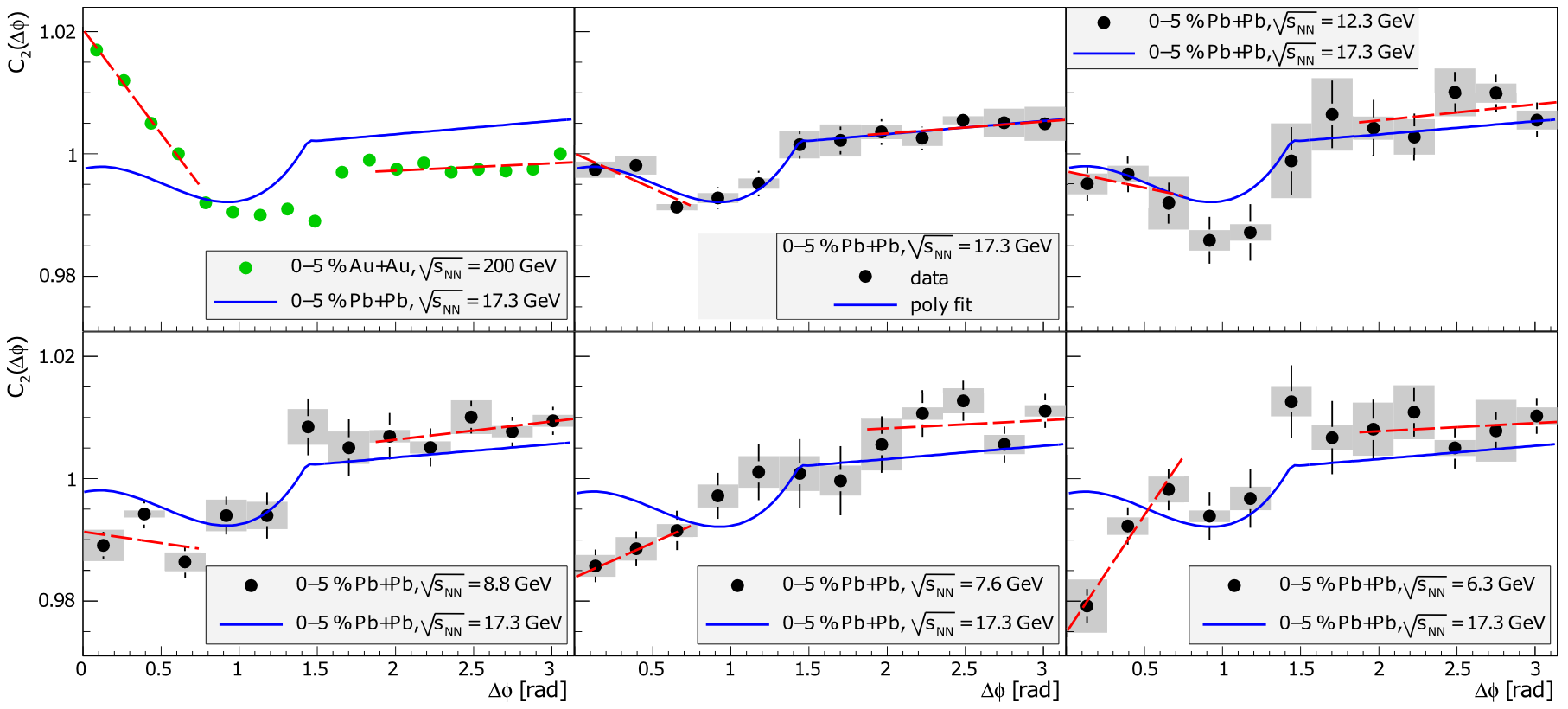}
  \end{center}
  \caption{Two-particle correlation functions from central \textit{Pb+Pb} (\textit{Au+Au}) events at $\sqrt{s_{NN}}$ = 
    200, 12.3, 8.8, 7.6 and 6,3~GeV compared to results from central \textit{Pb+Pb} collisions at 17.3~GeV. Dashed
    lines illustrate linear fits used to extract slope parameters, plotted in Figure~\ref{fig:urqmd_Escan}.}
  \label{fig:energyScan}
\end{figure}

In Figure~\ref{fig:energyScan} the correlation function from central \textit{Pb+Pb} collisions
at $\sqrt{s_{NN}}$ = 17.3~GeV is compared to results from the same system at 12.3, 8.8, 7.6 and 6.3~GeV;
for illustration we also included a function obtained by PHENIX at the RHIC (\textit{Au+Au} collisions at $\sqrt{s_{NN}}$ = 200~GeV),
for the same centrality and $p_{T}$ ranges~\cite{phenixFcn}. Dashed lines depict linear fits of the near and away sides
of $C_{2}(\Delta\phi)$. The resulting slope parameters are plotted in Figure~\ref{fig:urqmd_Escan}. The near-side peak appears
to turn into depletion with decreasing energy, whereas shape and amplitude of the away-side enhancement remains mostly unchanged
throughout the scan. Should the flattening of the latter be considered a quark-gluon plasma signature, these results are at odds with present-day
expectations that the QGP is produced only at higher energies.

\subsection{Comparison with UrQMD}
\label{sec:comparisonWithUrQMD}

\begin{figure}[htb]
  \begin{center}
    \includegraphics[width=0.49\textwidth]{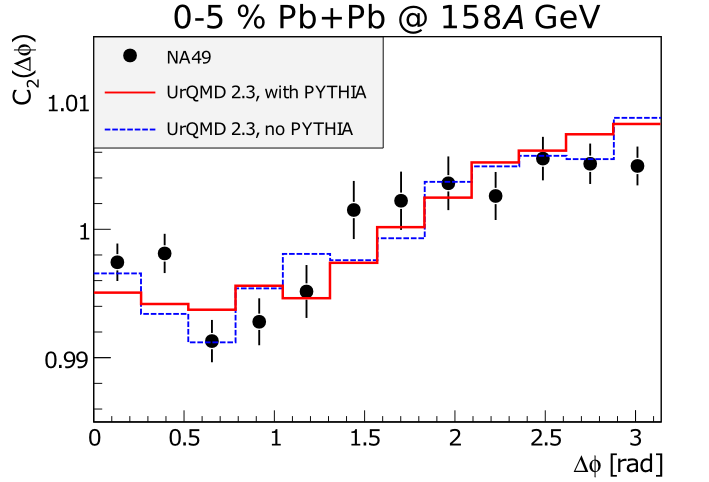}
    \includegraphics[width=0.49\textwidth]{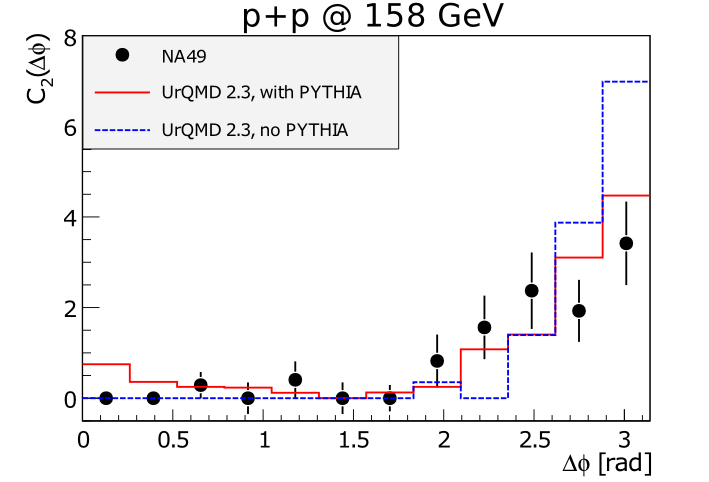}
  \end{center}
  \caption{Comparison of experimental and simulated correlation functions for central \textit{Pb+Pb} (left) and \textit{p+p}
    (right) collisions at $\sqrt{s_{NN}}$ = 17.3~GeV. Points: experimental data, solid lines: UrQMD with PYTHIA,
    dashed lines: UrQMD without PYTHIA. Systematic errors have been omitted for clarity.}
  \label{fig:urqmd_Pb_p}
\end{figure}

A comparison of real-data azimuthal correlation functions from central-\textit{Pb+Pb} and \textit{p+p} collisions
at $\sqrt{s_{NN}}$ = 17.3~GeV with results obtained from UrQMD simulations can be found in Figure~\ref{fig:urqmd_Pb_p}.
For both systems good agreement can be observed between the data and the simulations, especially on the away side.
Moreover, strong similarity of correlation functions with and without jet contribution imply this particular
correlation source not to play a major role in the SPS energy range.

\begin{figure}[htb]
  \begin{center}
    \includegraphics[width=0.49\textwidth]{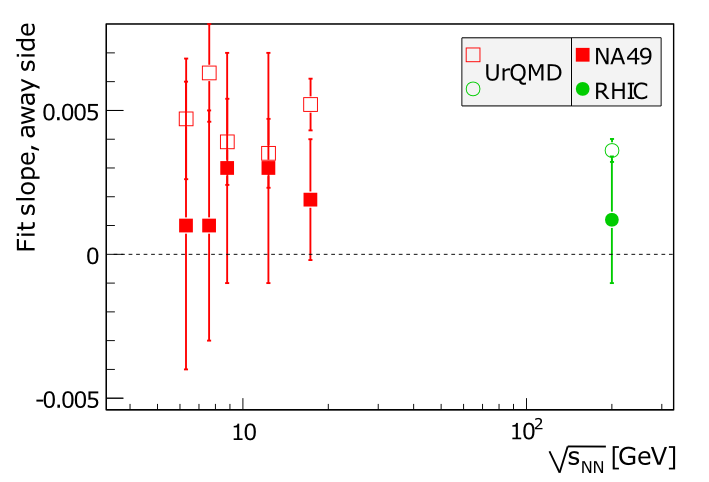}
    \includegraphics[width=0.49\textwidth]{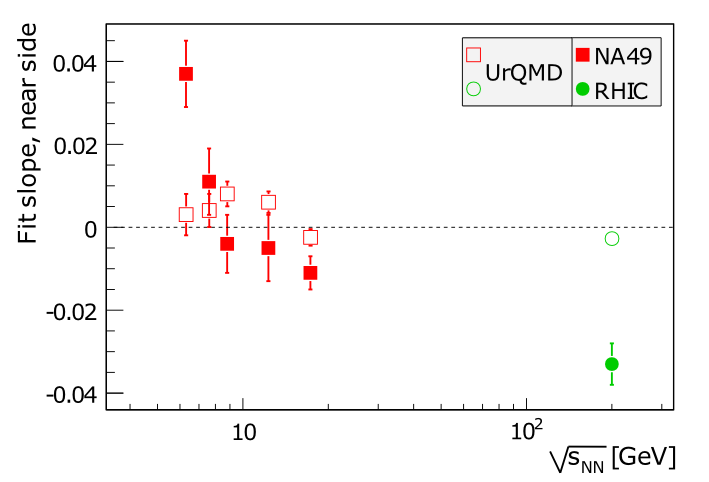}
  \end{center}
  \caption{Dependence of slope values extracted from real-data (full points) and UrQMD (open points) correlation
    functions as a function of collision energy. Left: away side, right: near side.}
  \label{fig:urqmd_Escan}
\end{figure}

Correlation functions for lower-energy UrQMD data sets were also produced but are not shown due
to limited space. Slopes of linear fits to both UrQMD simulations and real data are shown in Figure~\ref{fig:urqmd_Escan}.
It can clearly be seen here that the weak dependence of away-side slope on energy in real data is also present in UrQMD
simulations. However, data and simulations follow different trends on the near side.

\section{Summary}
\label{sec:summary}

A system-size and energy scan of two-particle correlation functions at high $p_{T}$ was performed by NA49.
A flattening of the function's away side was observed in central heavy-ion collisions even
at low SPS energies, raising doubts about the standard parton energy loss interpretation. Interestingly,
UrQMD predictions agree well with the away-side experimental results.

On the other hand, clear energy dependence of the correlation function was observed
on the near side. As the energy decreases an enhancement changes into a depletion in the region
close to the probable onset of deconfinement. Further studies are required to find out whether
the coincidence of these two features is related or accidental.

 % do not change 

\end{document}